\documentstyle[epsf]{mn}

\title[{\sl XMM-Newton} observations of Her X-1 over the
35~day beat period]{{\sl XMM-Newton}
EPIC \& OM observations of Her X-1 over the
35~day beat period}

\author[]{Silvia Zane$^{1}$, Gavin Ramsay$^{1}$, Mario A.
Jimenez-Garate$^{2}$, Jan Willem den Herder$^{3}$,\and 
Charles J. Hailey$^{4}$ 
\\
$^{1}$Mullard Space Science Laboratory, University College London,
Holmbury St. Mary, Dorking, Surrey, RH5 6NT, UK\\
$^{2}$MIT Center for Space Research, 77 Massachusetts Avenue, Cambridge,
MA 02139, USA\\
$^{3}$SRON, the National Institute for Space Research,
Sorbonnelaan 2,
   3584 CA Utrecht, The Netherlands\\
$^{4}$Columbia Astrophysics Laboratory, Columbia University, New York,
   NY 10027, USA\\}
\date{Accepted 23 Jan 2004: }

\begin{document}
\def\rchi{{${\chi}_{\nu}^{2}$}}
\def\uchi{{${\chi}^{2}$}}
\outer\def\gtae {$\buildrel {\lower3pt\hbox{$>$}} \over
{\lower2pt\hbox{$\sim$}} $}
\outer\def\ltae {$\buildrel {\lower3pt\hbox{$<$}} \over
{\lower2pt\hbox{$\sim$}} $}
\newcommand{\Msun} {$M_{\odot}$}
\newcommand{\xmm} {\sl XMM-Newton}

\maketitle

\begin{abstract}

We present the results of a series of {\xmm} EPIC and OM observations of
Her X-1, spread over a wide range of the 35~day precession period. We 
confirm 
that the spin modulation of the neutron star is weak or absent in the low
state - in marked contrast to the main or short-on states. 
During the states of higher intensity, we observe a
substructure in the broad soft X-ray modulation below $\sim 1$~keV,
revealing the presence of separate peaks which reflect the structure seen 
at higher 
energies. The strong fluorescence emission line at
$\sim$6.4~keV is detected in all observations (apart from one taken in the
middle of eclipse), with higher line energy, width and normalisation
during the main-on state. In addition, we report the detection of a second
line near 7~keV in 10 of the 15 observations taken during the
low-intensity states of the system. This feature is rather weak and not
significantly detected during the main-on state, when the strong continuum 
emission
dominates the X-ray spectrum. Spin resolved spectroscopy just after the
rise to the main-on state shows that the variation of the Fe K$\alpha$ at 
6.4 keV is correlated with the soft X-ray emission. This confirms our
past finding based on the {\xmm} observations made further into the
main-on state, and indicates the common origin for the thermal component 
and the Fe K$\alpha$ line 
detected at these phases. We also find that the normalisation of the 6.4keV line 
during
the low state is correlated with the binary orbital phase, having a broad
maximum centered near $\phi_{orbit}\sim0.5$. We discuss these observations
in the context of previous observations, investigate the origin of the
soft and hard X-rays and consider the emission site of the 6.4keV and 7keV
emission lines.

\end{abstract}

\begin{keywords} accretion, accretion disks -- X-rays: binaries --
individual: Her X-1 -- stars: neutron \end{keywords}

\section{Introduction}
\label{int}

Her X-1 is one of the best studied X-ray binaries in the sky; this is
borne out by the fact that since 2000 there has been more than 30
papers published in refereed journals with ``Her X-1'' in the
title. We list very briefly the most salient observational
characteristics which are relevant to the present study. Her X-1 is a
binary system, which consists of a neutron star and an A/F secondary
star; it has an orbital period of 1.7~day and the spin period of the
neutron star is $\sim$1.24~sec. It is one of only a few systems which
shows a regular variation in X-rays: the 35 day variation is referred
to as the ``beat'' period throughout this paper, and the phasing on
this period is marked as $\Phi_{35}$.

The beat cycle begins with a sharp rise in X-ray intensity, which reach
the maximum flux, $F_{max}$, in less than $\sim$0.5~day and then lasts
$\sim$10~day (``main-on'' state). This is followed by a period of low
X-ray emission (``low-state'') and later again by a ``short-on'' phase,
during which the X-ray flux reaches $\sim 1/20 F_{max}$ and $\sim 1/3
F_{max}$, respectively.  A second low-state follows until the onset of the
next main-on. The spin modulation of the neutron star is most prominent
during the main-on state, becoming much less so (if seen at all) during
the low-state and increasing again during the short-on (see e.g. Deeter et
al.~1998).

The observed X-ray spectrum changes markedly over the beat
cycle. However, it has been recognized that the X-ray spectra taken at
different beat phases can be fitted using a constant emission model
and adding one or more absorption components which vary in column
density over the beat period, reaching a maximum value during the
low-state (e.g. Mihara et al.~1991, Leahy~2001).  This strengthened
the idea that the 35~day cycle is due to the precession of an
accretion disk, that periodically obscures the neutron star beam.

A rather broad Fe K$\alpha$ emission line at $\sim$6.5~keV has been
observed by {\sl Ginga} (see Leahy, 2001 and past references therein);
line energy and width have been found to vary over both the beat and
the spin period (e.g. Choi et al.~1994). Using {\sl ASCA} data taken
during the main-on state, Endo, Nagase \& Mihara (2000) were able to
resolve the feature into two discrete emission lines, at $\sim$6.4~keV
and 6.7~keV.  However, the second feature has never been detected in
other observations.  With the advent of imaging X-ray satellites with
large effective area over a broad energy band and relatively good
spectral resolution at energies near 6~keV, it is now possible to
resolve the Fe lines with better signal to noise than what was
possible using {\sl ASCA}. Moreover, while {\sl ASCA} had a lower
limit of $\sim$0.5~keV, the energy response of {\xmm} extends down to
$\sim$0.2~keV and this allows a more accurate quantification of the
absorption column density.

Another manifestation of the 35~day cycle is the evolution of the
Her~X-1 pulse profile at energies above $\sim 1$~keV. A few other
pulsars show pulse shape changes which are correlated with high-low
states of X-ray intensity, but they are typically transients such as
GX~1+4 (Dotani et al.~1984) and EXO~2030+375 (Parmar, White \&
Stella~1989) or flaring sources as LMC~X-4 (Levine et
al.~1991). Her~X-1 is unique in having a repetitive evolutionary
pattern in pulse changes tied to a regular high-low intensity cycle,
and is therefore a subject of continued interest.  Particularly
successful in explaining the basic features observed by {\sl Ginga}
and {\sl RXTE} is the accretion column model by Scott et
al.~(2000). This model invokes the successive obscuration of a compact
pencil beam from the neutron star, two fan beams with size $\sim
2R_{ns}$ and a more extended scattered component.

At lower energies (below $\sim 0.7$~keV), the light curve does not
show evidence for a similar variety of emission components. Past
detections of pulsations in the soft X-rays of Her~X-1 obtained using
{\sl EXOSAT} (\"{O}gelmen \& Tr\"{u}mper~1988), {\sl ROSAT}
(Mavromatakis~1993), and {\sl BeppoSax} (Oosterbroek et al.~1997,
2000, 2001) only show a broad and quasi-sinusoidal modulation, which
is shifted in phase with respect to the maxima observed at higher
energies. The phase shift, as measured during the main-on and short-on
state, differs from $180^\circ$. Under the assumption that soft X-rays
are due to the reprocessing of the pulsar beam by the inner edge of
the disk, this is usually interpreted as evidence for a tilt angle in
the disk (see e.g. Oosterbroek at al.~1997, 2000).  As predicted by
precessing disk models (Gerend \& Boynton, 1976), the hard/soft shift
in phase should vary along the beat cycle. The ideal way to test this
would be to track the pulse difference over the entire 35~day
period. However, past measurements were restricted to the main-on and
short-on states, due to the difficulty of detecting a spin modulation
during phases in which the system is highly absorbed. Using {\xmm}\/
data, Ramsay et al.~(2002, hereafter R02) reported the first evidence
for a substantial change in the phase difference along the beat
cycle. We also found that the values of the phase shift observed by
{\xmm}\/ during the main-on and short-on differ from those measured in
the past at similar beat phases. However, at the time of that paper
only three {\xmm}\/ observations, at three different beat phases, had been 
performed.

To date, Her X-1 has been observed by {\xmm} on 15 separate occasions
giving good coverage over the beat period. The first three
observations have been discussed in detail by R02 and Jimenez-Garate
et al.~(2002).  Here we report the analysis of the remaining datasets,
focusing on results obtained using the broad energy band EPIC
detectors and putting our past findings in a broader contest.  We also
present the simultaneous UV observations made using the Optical
Monitor (OM). Observations are described in \S~\ref{obs}. UV data and 
timing analysis are presented in \S~\ref{uvdata} and \S~\ref{spin} 
respectively, while the behaviour of the Fe feature(s) over the beat, the 
spin and the 
ornital period is reported in \S~\ref{spec}. Summary and
discussion follow in \S~\ref{disc}.

\section{Observations}
\label{obs}

\subsection{Instrumental setup and data reduction}

{\xmm} has three broad-band imaging detectors with moderate spectral
resolution: two EPIC~MOS detectors (Turner et al.~2001) and one EPIC~PN
detector (Str\"{u}der et al.~2001). During the observations of Her~X-1, PN
and MOS1 were configurated in timing mode,
while MOS2 was configured in
either small window or full frame mode. All three EPIC cameras were
used in conjunction with medium filters.
The Optical Monitor (Mason et al.~2001) was
configured in imaging mode, by using the UVW1 and UVW2 filters
(effective wavelength 2910 and 2120 \AA\hspace{1mm} respectively).

The details of the observations are summarized in Table~\ref{obslog};
as previously mentioned, the results from the first three observations
have been already reported in detail by R02 and Jimenez-Garate et
al.~(2002).  The raw event files were first examined and corrected for
any time anomalies (see R02), and then reduced using the {\xmm}
Science Analysis Software v5.3.3. The events from timing mode data
were extracted in a way similar as in R02.  For datasets taken in
full-frame and small window mode, events were extracted from an
aperture of $\sim 40''$ centered on the source and a background,
source-free region; the background subtracted spectra were obtained by
scaling appropriately for the size of the two regions. Finally, the
event arrival times were corrected to the solar system barycenter and
to the center of the binary system, using the ephemeris of Still et
al.~(2001).

\subsection{Determining the beat phase}

\begin{figure}
\begin{center}
\setlength{\unitlength}{1cm}
\begin{picture}(8,9)
\put(-0.1,-2){\includegraphics{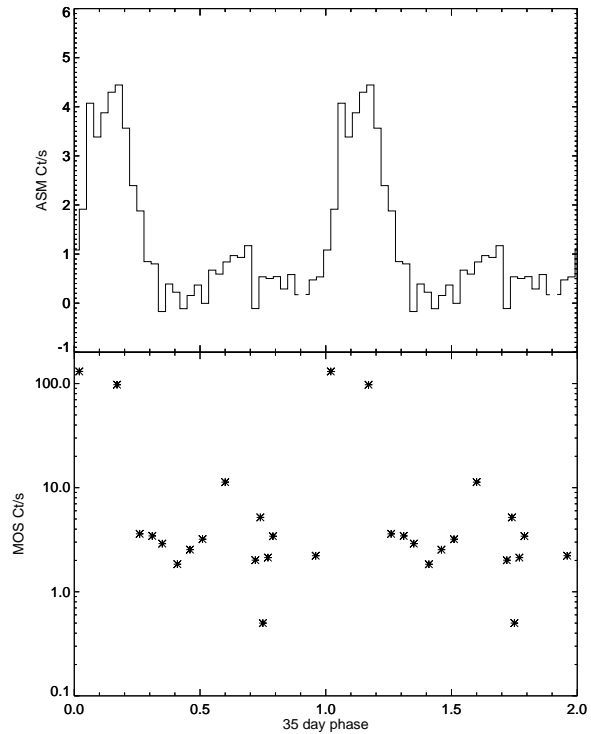}}
\end{picture}
\end{center}
\caption{Top Panel: a typical {\sl RXTE/ASM} (2-10)~keV light curve of
Her~X-1 over the 35 day beat period. Bottom Panel: the mean count rate
of Her~X-1 registered by the EPIC MOS detectors. The lowest count rate
corresponds to an eclipse of the neutron star by the secondary.}
\label{asm}
\end{figure}

The beat period of Her~X-1 is known
to vary by up to a few days (Baykal et al.~1993).
In order to determine how the epochs of the {\xmm} observations relate to
the precession cycle, we
extracted the {\sl RXTE ASM} (2-10)~keV quick-look light curve from
the MIT web site and we
determined the 35~day phase by computing the
time difference between each observation and the start of the previous
rise to maximum. We then took the period to be the time till the
next rise to maximum. We estimate the uncertainty on the
phase of the beat period to be $\sim$0.01-0.02 cycles. The resulting beat
phase is reported in Table~\ref{obslog}, along with the dates of the
observations and the mean X-ray brightness.
Note that the last observation has been taken during an eclipse of the
primary by the secondary star: this corresponds to the point with the lowest
X-ray count
rate in Figure~\ref{asm}.

\begin{table}
\begin{tabular}{lllllr}
\hline
Observation& {\sl XMM-} & Exp&MOS&Orbital&35~d\\
Date & {\sl Newton} & (ksec) &Mean&Phase&Phase\\
     & Orbit & & Ct/s& \\
\hline
2001 Jan 26 & 207  & 10& 97.5 & 0.20--0.26&0.17\\
2001 Mar 04  & 226  & 11& 3.6 & 0.47--0.56&0.26\\
2001 Mar 17 & 232  & 11& 11.3 & 0.52--0.60&0.60\\
2002 Feb 25 & 405 & 10.3  & 2.5 & 0.55-0.62 & 0.46 \\
2002 Feb 27 & 406 & 11.7 & 3.2 & 0.71-0.79 & 0.51 \\
2002 Mar 07 & 410 & 8.8 & 5.2 & 0.40-0.46 & 0.74 \\
2002 Mar 09 & 411 & 7.3 & 3.4 & 0.57-0.62 & 0.79 \\
2002 Mar 15 & 414 & 7.3 & 2.2 & 0.09-0.14 & 0.96 \\
2002 Mar 17 & 415 & 7.3 & 130 & 0.26-0.31 & 0.02 \\
2002 Mar 27 & 420 & 4.4 & 3.4 & 0.17-0.20 & 0.31 \\
2002 Sep 20 & 509 & 5.9 & 2.9 & 0.56-0.60 & 0.35 \\
2002 Sep 22 & 510 & 7.3 & 1.8 & 0.72-0.77 & 0.41 \\
2002 Oct 04 & 516 & 10.3 & 2.1 & 0.78-0.85 & 0.77 \\
2003 Mar 26 & 603 & 15.6 & 2.0 & 0.81-0.88 & 0.72 \\
2003 Mar 28& 604 & 6.1 & 0.5 & 0.98-0.03 & 0.75 \\
\hline
\end{tabular}
\caption{Summary of the {\sl XMM-Newton} observations of Her X-1. For
the orbital phasing we use the ephemeris of Still et al.~(2001); for
the 35~d phasing we define phase 0.0 at the main high state turn-on (see
text for a discussion). The count rate refers to the rate in either the
MOS1 or MOS2 detector depending on the source intensity. 
Analysis of the first three datasets has been
presented in details in R02 and Jimenez-Garate et
al.~(2002).} \label{obslog}
\end{table}

The phase coverage of the {\xmm} observations over the 35~day period is
shown in Figure~\ref{asm}. For comparison, we also show the ASM 
light curve recorded during a typical cycle. We 
note that, since 
the actual period of the X-ray modulation varies around its nominal 
value of 35~d, the determination of the state of the system (either 
low, main on or short on) based on a direct comparison with the ASM 
lightcurve in the top panel must be taken only as indicative. 

As we can
see, the beat cycle is reasonably well sampled, with most of the exposures
taken outside the main-on and short-on phases. Three observations have
been performed during states of high intensity: in addition to the main-on
and short-on datasets discussed by R02 ($\Phi_{35}=0.17$ and 0.60)
we have a further exposure at $\Phi_{35}=0.02$. We note that the ASM 
lightcurve

\section{The UV data}
\label{uvdata}

\begin{figure}
\begin{center}
\setlength{\unitlength}{1cm}
\begin{picture}(8,9)
\put(-0.1,-2){\includegraphics{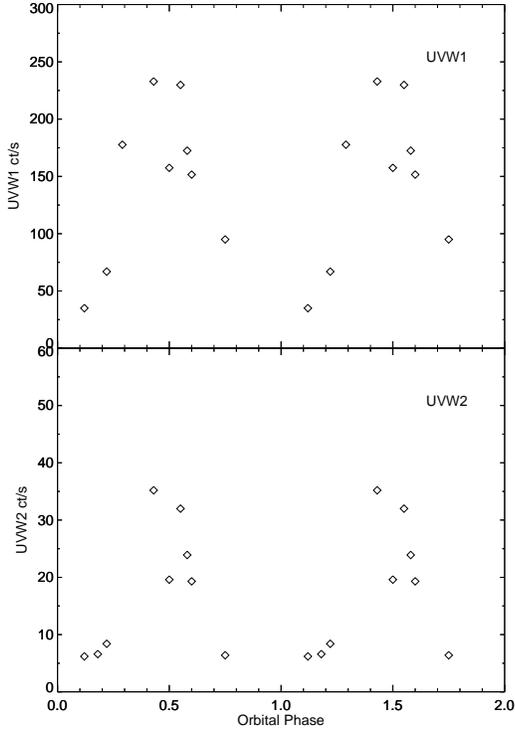}}
\end{picture}
\end{center}
\caption{Top Panel:
The UVW1 (top panel) and UVW2 (bottom panel) data folded on the orbital
period. data have been phased by using the
ephemeris of Still et al.~(2001).}
\label{uv}
\end{figure}

The UV emission probably originates from a blend of at least two
different components: the illuminated face of the secondary star
and the accretion disk. Their relative contribution varies during
the orbital motion, and in general is difficult to disentangle.
Unlike the X-rays, the UV emission does not show an obvious
correlation with the beat period. This is in agreement with the
fact that the disk contribution is on average small (less than
10\% as estimated by Boroson et al.~2001). However, it is
modulated over the binary orbital period. The UV count rate
detected using the two OM filters is shown in Figure~\ref{uv}, as
a function of the orbital phase. Note that observations were not
always performed in both filters, and in the last two exposures no
UV measurements were made. The illuminated secondary star is
expected to dominate around $\phi_{binary}\sim0.5$. As we can see,
near this orbital phase the UV flux measured by OM reaches a broad
maximum, similar to that found using $IUE$ in the
1750-1850\AA\hspace{1mm} band (e.g. Vrtilek \& Cheng~1996).
However, there is a flux decrease at $\phi_{binary}\sim$0.5,
possibly due to the fact that part of the secondary star is
shadowed by the accretion disk or to a warp in the disk itself.

The rise from the
eclipse ($\phi_{binary}\sim$0.1-0.3) is steeper in UVW1 (longer
wavelength) than
in UVW2. Since, at these orbital phases, the accretion disk mainly contributes to the UV flux, this
is consistent with a scenario in which,
after the eclipse, the regions of the disk farthest from the neutron
star come into view first.

\section{Timing analysis}
\label{spin}

\subsection{Search for modulation close to the spin period}

The spin period of Her X-1 has been found to vary from $\sim$1.23772~s
to 1.23782~s, with alternating phases of spin-up and spin-down (Parmar
et al. 1999, Oosterbroek at al. 2001). Furthermore, it can vary
significantly on relatively short time scales.
The values of the spin period measured using the first three {\xmm}
observations were reported by R02: using the PN
detector, we found 1.237774~s, 1.237753~s and 1.237751~s, respectively.
The first measure confirmed the slow-down trend monitored by
{\sl BeppoSax} (Oosterbroek et al.~2000) and Chandra (Burwitz, private
communication); although we observed a slight decrease in the two
following epochs.

We performed a search for pulsations using a Discrete Fourier
Transform for all the new datasets, finding a strong evidence of
pulsations only in one case (revolution 415). This is the only new
exposure taken outside the low state, precisely during the main-on at
$\Phi_{35}=0.02$. The corresponding spin period, as obtained from the
EPIC PN data, is 1.237777(1)~s, where the number in parenthesis
represents the error on the last digit as determined by using the
Cash~(1979) statistic. The spin period is just slightly increased with
respect to the value measured by {\xmm} in 2001.

\begin{figure}
\begin{center}
\setlength{\unitlength}{1cm}
\begin{picture}(8,9)
\put(-0.7,-1.3){\includegraphics{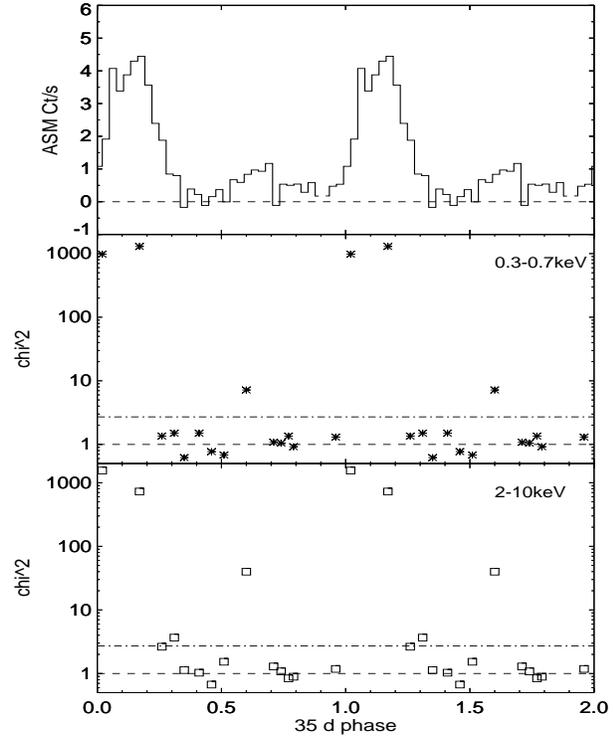}}
\end{picture}
\end{center}
\caption{Top panel: the mean ASM X-ray light curve over the 35~day
cycle. Middle panel: the \uchi obtained by fitting a constant 
value to the 0.3-0.7~keV folded light curve. Bottom panel: same but for 
2-10~keV
folded light curve. In folding the data we folded the data on the spin
period determined using a Discrete Fourier Transform (main-on or
short-on observations) or, when unavailable, the more recent measure
of 1.237777~s.  The \uchi is computed by using 25 phase bins.}
\label{spin_var}
\end{figure}

All the remaining datasets showed amplitude spectra with only weak or
no significant peaks close to the expected spin period.  In order to
investigate further the possible presence of a modulation, we folded
all the datasets on the most recent measured period, i.e. 1.237777~s,
by using 25 phase bins. We then fitted a constant value to the
resulting light curve and determined the \uchi. Therefore, we expect a
poorer fit in datasets with a stronger spin modulation. Since the
profile of the folded light curve varies considerably in the soft 
and
hard energy band, we split the events into two energy ranges:
0.3-0.7~keV and 2-10~keV.  The resulting \uchi\/ is shown in
Figure~\ref{spin_var}, as a function of the 35~day phase.  For 1
degree of freedom (the value of the period), the difference to the fit
is $\Delta$\uchi=2.71 at the 90 percent confidence level.

Using this method we find that, outside the main-on and short-on
states there is evidence for significant modulation above 2~keV in the
two datasets taken at $\Phi_{35}$=0.26 and 0.31. In the first case
(which is one of those already reported by R02) the \uchi\hspace{1mm}
increases above the 90 percent confidence level if we reduce the
number of phase bins from 25 to 20.

Defining the amplitude as (max-min/mean) of a fitted sinusoid, we find
an amplitude of 11 and 13 percent for $\Phi_{35}$=0.26 and 0.31,
respectively.  Previous observations of Her X-1 during the low state
include Coburn et al.~(2000), who found a spin modulation of $\sim$13
percent (assuming the same definition as ours) in the 3--18~keV range
using {\sl RXTE} data in an anomalous low state, while Mihara et
al.~(1991) determined an upper limit for the pulsed fraction of 2.4
percent in the 1.2-37~keV energy range using {\sl Ginga} data.

\begin{figure*} \begin{center} \setlength{\unitlength}{1cm}
\begin{picture}(8,10) \put(14.3,0.4){\includegraphics{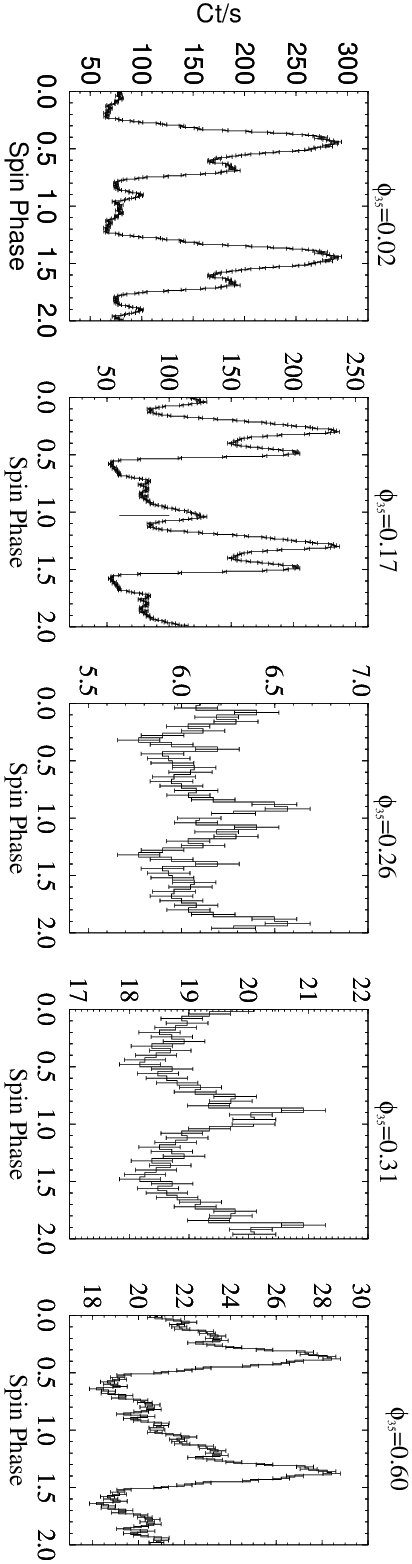}}
\put(14.3,-5){\includegraphics{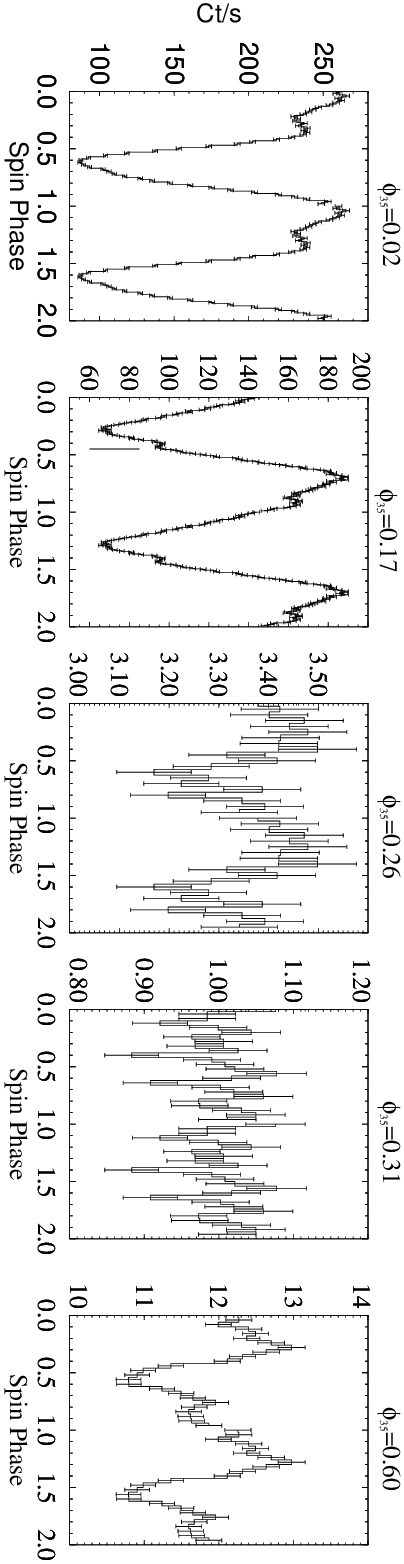}} \end{picture} \end{center} \caption{Upper
panel: the EPIC~PN spin profile in the 2-10~keV band as a function of the
35~day cycle. Lower panel: as above but for the 0.3-0.7 keV band.
In all panels, because of the uncertainty in the spin period in
each epoch, the spin phases are not on an absolute scale. The two
vertical lines at $\Phi_{35}$ = 0.17 mark the second atypical peak
in the interpulse (hard band) and the small ``notch'' like feature
(soft band) discussed in the text.} \label{spin_hard}
\end{figure*}

\subsection{Spin-resolved light curves}

In Figure \ref{spin_hard}, we show the spin profiles of Her X-1 as derived
for those observations in which we detected a significant modulation in
the 0.3-0.7~keV or in the 2-10~keV energy band. Due to the uncertainty in
the spin period at each epoch, the spin phases are not on the same
absolute scale. Therefore, only the relative phasing between the
light curves in different energy bands {\it taken during the same
observation} is physically meaningful (see \S~\ref{phase_shift}).

The change of the spin profile of Her~X-1 in different energy bands
has been well documented in the literature. In order to compare the
spin profiles obtained by using {\xmm} with the previously published
data, we take as reference, in the range above $\sim 2$~keV, the work
of Deeter et al.~(1998) and Scott et al.~(2000). These authors
reported the variation of the hard X-rays light curve obtained by {\sl
Ginga} and {\sl RXTE} over a wide range of beat phases.

As we can see, the {\xmm} light curves obtained shortly after the
main-on turn on ($\Phi_{35}$=0.02) in the 2-10~keV energy band closely
resemble most of the main-on states previously reported. The only
exception is that the small peak observed during the inter-pulse (at
$\phi_{spin}$=0.9 in Figure \ref{spin_hard}) appears slightly more
sharp than previously reported. At the next main-on beat period,
$\Phi_{35}$=0.17, we detect an atypical profile, with a second large
peak in the interpulse on the leading side of the main pulse (the
feature occurs at $\phi_{spin}$=0.05 in Figure \ref{spin_hard}).  The
peak becomes more prominent as the high state progresses, and has been
observed only occasionally in Her~X-1: it is present in the {\it
Ginga} data taken during the main-on state ``D'' by Deeter et
al.~(1998), at the two beat phases $\Phi_{35}$=0.073 and
0.162. However, it was not present in the other of their main-on
observations nor in those discussed by Scott et al.~(2000). Deeter et
al.~(1998) noted that such atypical profile is similar to a previous
one recorded by using {\sl Kvant} (Sunyaev et al.~1988) and suggest
it may be due to a reduction in the accretion rate.

Moving later in beat phase, the detected amplitude of the hard X-ray
modulation decreases sharply at $\Phi_{35}$=0.26 and $\Phi_{35}$=0.31:
here the modulation is broad and quasi-sinusoidal. There is evidence
of a double peaked structure at $\Phi_{35}$=0.26: R02 show the hard
X-ray modulation resolved into finer energy bands and find that at
this beat phase modulation at the hardest energies is broad and
single-peaked. Note that the profile obtained at $\Phi_{35}$=0.31 is
less certain, since it has been obtained by folding data over the most
recent measured period and not using the period measured at the time
of the observation. However, we have explored the effect of folding
the data on slightly different spin periods, centered around
1.237777~s: the resulting folded light curves are very similar.  The
profile detected during the short-on ($\Phi_{35}$=0.60) is similar to
previous observations at a similar beat phase, although the high
signal to noise ratio of the {\xmm} data revealed a more complex
structure during the rise of the main peak.

The high throughput of {\xmm}\/ allows us to obtain a light curve with
very high signal to noise ratio also in the soft X-rays ($0.3-$1~keV).
Consistent with past measurements, the spin profiles of the soft X-ray
band in Figure~\ref{spin_hard} show a basically sinusoidal shape.
However, in both main-on observations the broad peak is cut by a dip
which has not be seen before. Further, data taken at $\Phi_{35}$=0.17.
show a small ``notch'' like feature preceding the maximum (at
$\phi_{spin}$=0.4 in Figure~\ref{spin_hard}). During the short-on
($\Phi_{35}$=0.60) the modulation is broader, with evidence for a
fairly complex substructure during the leading edge of the peak.

\subsection{The phase shift between the soft and hard light curves}
\label{phase_shift}

R02 found that the phase shift between the soft and hard X-ray light
curves, folded on the spin period, varied over the beat period. To
determine how the dataset taken at $\Phi_{35}$=0.02 compared with
these previous observations we cross-correlated the hard and soft
light curves (see R02 for details). This is the only new observation
that can be used for such purpose, since the others do not show a
significant modulation in both soft and hard X-rays. As we can see,
the two observations taken during the main-on have a very similar
correlation, with data taken at $\Phi_{35}$=0.02 showing a strong
anti-correlation and a phase shift between the maxima of the soft and
hard light curves of $\sim130^{\circ}$. This difference in phase is
slightly smaller than what was observed at $\Phi_{35}$=0.17, and
consistent with the general trend over the beat period.

\begin{figure}
\begin{center}
\setlength{\unitlength}{1cm}
\begin{picture}(8,5.2)
\put(-0.8,-.7){\includegraphics{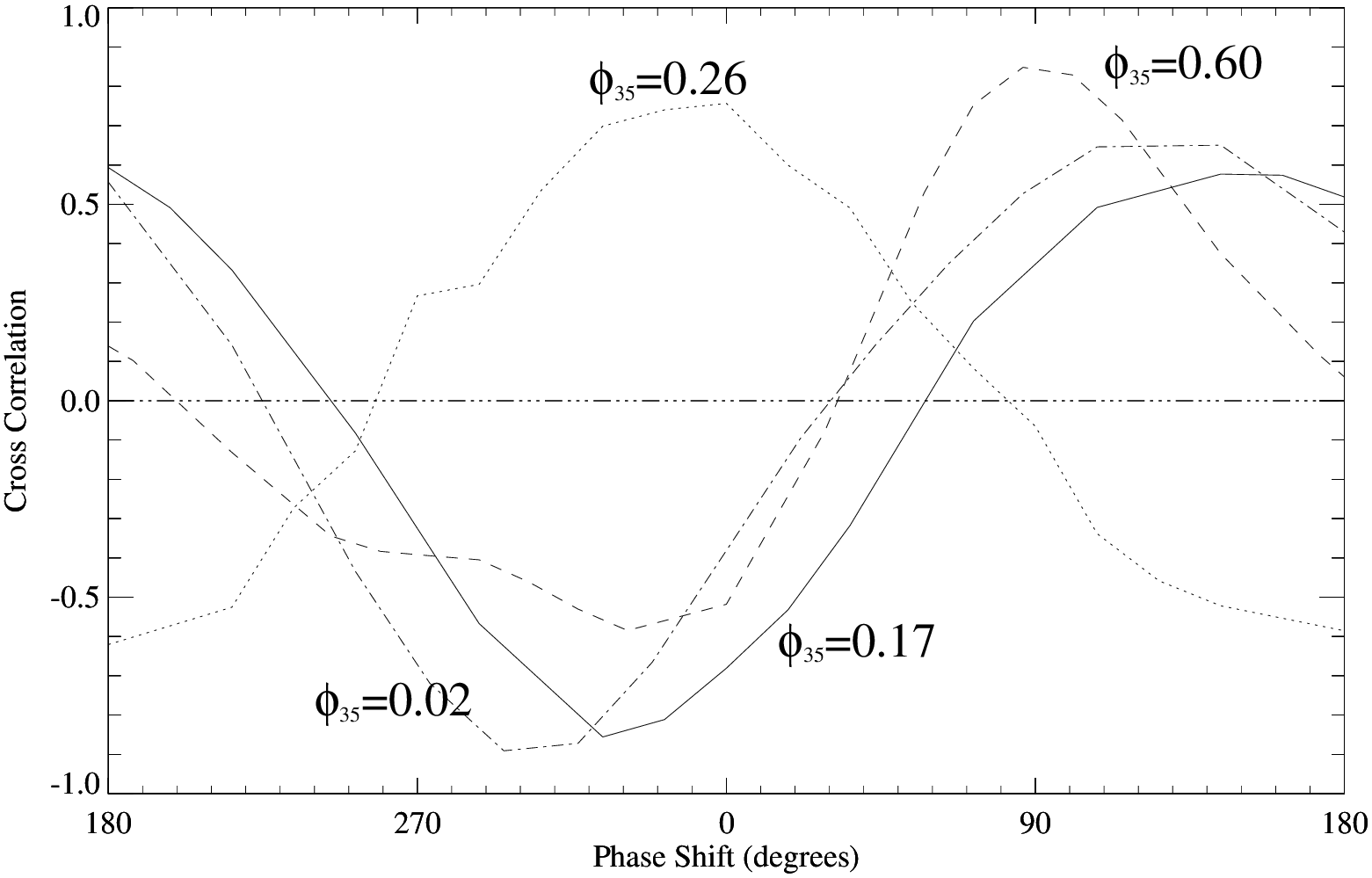}}
\end{picture}
\end{center}
\caption{The cross correlations for the soft (0.3-0.7~keV) and hard
(2-10~keV) light curves at four different 35~day phases.}
\label{cross_cor}
\end{figure}

\section{Spectral analysis: the variation of Fe complex}
\label{spec}

\subsection{Line variation over the beat cycle} 

\label{fe_35_var}

The integrated X-ray spectra of Her~X-1 over the 35 day cycle have
been modeled by many authors (e.g. Mihara et al.~1991, Oosterbroek at
al.  ~1997, Coburn et al.~2000, Oosterbroek at al.~2000, Choi et
al.~1994, Leahy 2001). In accordance with these studies, the main 
on, short on and low states all require a two component continuum with a
basic emission spectrum and an absorbed component whose strength is
maximum during the low states.  The natural interpretation is that the
35~day variation is due to the occultation of the neutron star beam by
the intervening accretion disk. Model fitting of the {\xmm} spectra
obtained using the first three datasets has been reported by R02 and
confirm this picture. Here, we used the same model for fitting the
newest datasets: since the overall fits do not show new
characteristics we do not discuss further the details of the overall
spectral shape.

\begin{figure}
\begin{center}
\setlength{\unitlength}{1cm}
\begin{picture}(8,13.3)
\put(-0.8,-1.){\includegraphics{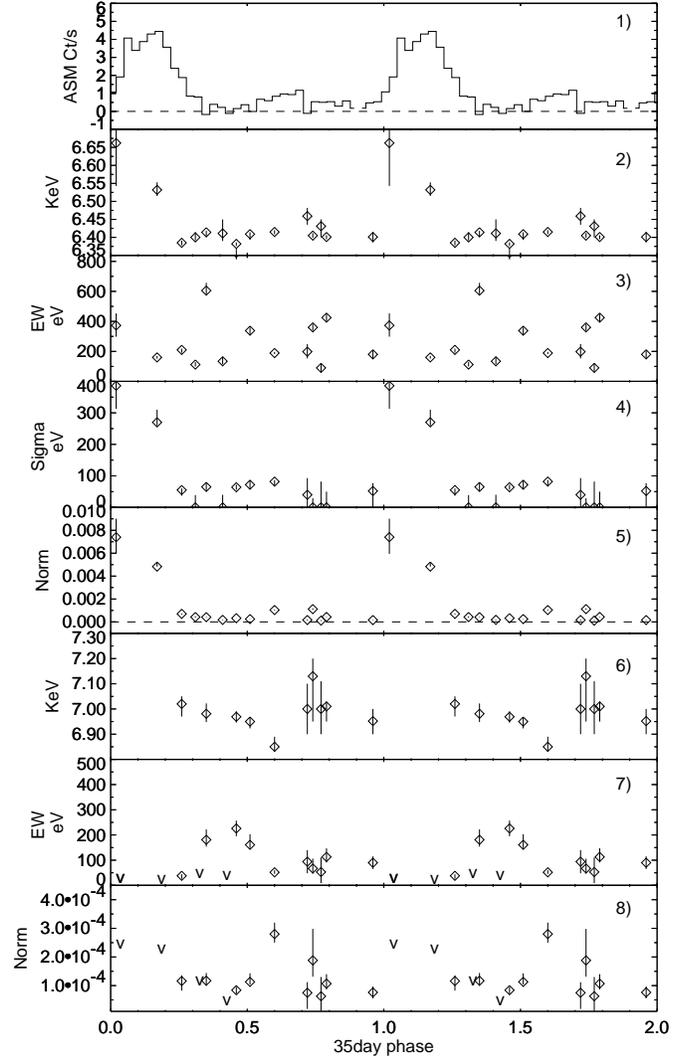}}
\end{picture}
\end{center}
\caption{Variation of the K$\alpha$ and Fe~XXVI line parameters along
the beat cycle, as measured with EPIC~PN. From the top: 1) mean ASM 
light curve; 2)-5) central
energy, equivalent width (EW), width and normalization of the
prominent K$\alpha$ Fe emission line; 6)-8) central energy, EW and
normalization of the Fe line at $\sim 7$~keV. For those datasets where
the second feature is undetected, we report upper limits to the EW and
normalization, at the 90 percent confidence level (``v'' symbols).}
\label{fe_line}
\end{figure}

Instead, we focus on the behaviour of the X-ray spectra near the Fe
line component(s), detected in the past around $\sim 6.6$~keV. As
shown by R02, the central energy of the fluorescent line varies over
the beat period, increasing during the main on. However, only three
epochs were reported in that paper.

In order to fit the spectra, we restricted the energy range to
4-10~keV and fitted an absorbed power law with one or more Gaussians.
The power law and Gaussian parameters were then fixed and the energy
range 0.3-4~keV included. The absorption component was allowed to vary
and we included a blackbody component and a neutral partial covering
model if required. This allows us to fix the absorption components at
an appropriate level, which is crucial since, for large values of the
column density, a characteristic absorption feature appears at $\sim
7.1$~keV. This feature may affect the identification of the emission
Fe lines, therefore it has to be modeled carefully.

Consistent with R02, we found that a prominent fluorescent line is
detected at all beat phases (apart from observation taken during
at mid-eclipse), with a central energy that varies between
$\sim$6.4-6.6~keV and is highest during the main-on. However, the most
striking result is the detection of a second component, an Fe XXVI
line at $\sim 7$~keV, which is only present outside the main-on. This
line was not reported by R02, but a closer analysis of the two
observations taken at $\Phi_{35}=0.26$ and 0.60 revealed its detection
in both datasets. The line parameters are reported, as a function of
the beat phase, in Figure~\ref{fe_line}.

We find that the equivalent width (EW) of the K$\alpha$ Fe line shows
a variation through the beat cycle, with no trend being visible.  This
is likely to be due to a varying continuum, since the K$\alpha$ Fe
line normalisation is highest during the main-on state while it 
remains low and almost
constant at other 35~day phases. Similarly, the line width is greatest
during the main-on state. The energy of the second component remains
approximately constant at $\sim$7.0~keV, except during the short-on
state where it is significantly lower ($\sim$6.85~keV).  The feature
is strongest (larger EW) during the states of low X-ray intensity,
although it is undetectable in some of the exposures in this phase
range. On the other hand, its normalisation is strongest during the
short-on. In some of the observations, we do not find evidence for a
significant Fe~XXVI line. In order to determine upper limits to EW and
line normalisation, we added a Gaussian component to the best fit
spectral model, by fixing the central energy at 7~keV and the width at
0.1~keV. The upper limits obtained for the datasets taken during the
main-on are consistent with the line being undetected because of the
strong continuum emission.

\begin{figure}
\begin{center}
\setlength{\unitlength}{1cm}
\begin{picture}(8,12)
\put(-0.2, 0.){\includegraphics{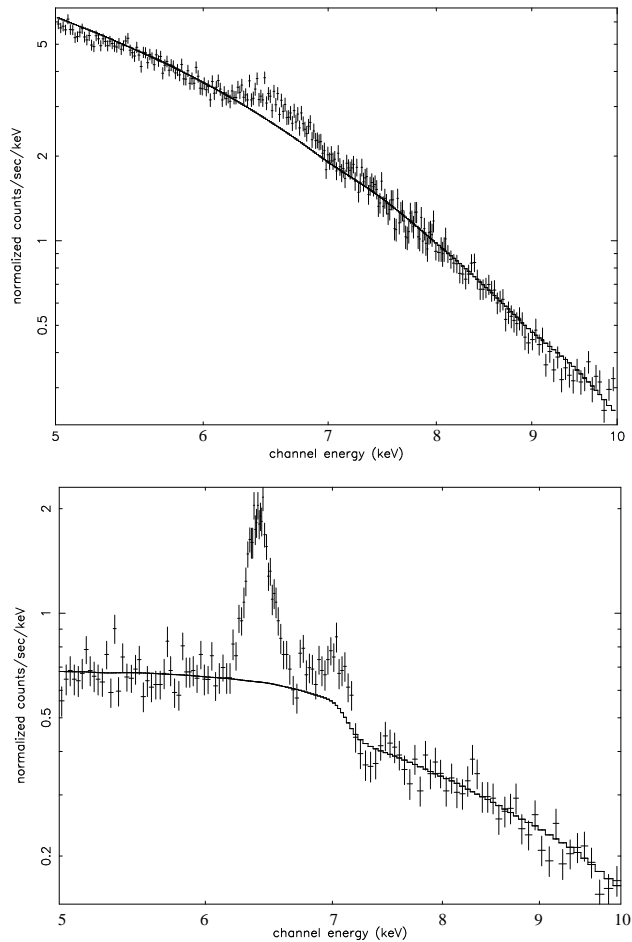}}
\end{picture}
\end{center}
\caption{The spectral region around 6.6keV from $\Phi_{35}$=0.02 (top) and
$\Phi_{35}$=0.79 (bottom). The y-axis is the normalised counts, i.e. 
the number of counts per energy bin per 
second, as measured by EPIC~PN.  
In both panel a solid line
shows the best fit after the normalisation of the one or two Gaussian
components have been set to zero.}
\label{fe_zoom}
\end{figure}

In order to show the significance of both the Fe features, we show in
Figure~\ref{fe_zoom} the X-ray spectrum between 5-9~keV for
$\Phi_{35}$=0.02 and $\Phi_{35}$=0.79. The solid line is the best fit
spectral model with the normalisations of the two Gaussian components
set to zero. We find no evidence for the presence of a
Compton downscattered shoulder (expected near $E_0 - E_0/(m_ec^2 + E_0)$ where 
$E_0$ is the 
centroid of the line) 
of the kind recently observed in HETGS {\sl Chandra} spectra 
of the high mass X-ray binary GX 301-2 (Watanabe et al.~2003). 

\subsection{Pulse-resolved spectroscopy}
\label{fe_spin}

The variation of the Fe K$\alpha$ line over the neutron star spin
period and its correlation with the thermal or non-thermal spectral
components was studied by R02. We found no evidence for a
significant variation in the line parameters at $\Phi_{35}$=0.26 or
0.60. In contrast, the observation made during the main-on
($\Phi_{35}$=0.17) showed a clear correlation between the EW of the
6.4keV line and the soft X-ray light curve. This supported the idea of
a common origin for the $\sim 6.4$~keV fluorescent line and the
thermal flux.

\begin{figure}
\begin{center}
\setlength{\unitlength}{1cm}
\begin{picture}(8,11.5)
\put(-0.8,-0.5){\includegraphics{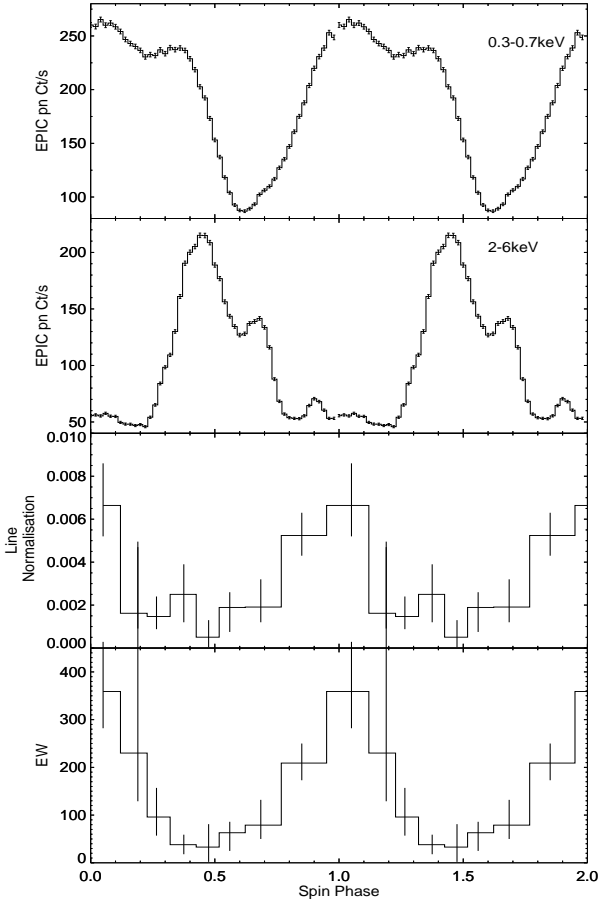}}
\end{picture}
\end{center}
\caption{The observation made using the EPIC PN detector at 
$\Phi_{35}$=0.02. From the top panel:
the light curve in the 0.3-0.7~keV band; the 2-6~keV band; the Fe
K$\alpha$ line normalisation and EW as a function of the spin
phase.}
\label{fe_line_correl}
\end{figure}

The new data taken during the main-on ($\Phi_{35}$=0.02) can be used
to test this scenario further. We phased all the events on the spin
period determined in \S\ref{spin} and extracted spin-phase resolved
spectra. Details of the model fitting are as in \S \ref{fe_35_var}.
Results are reported in Figure \ref{fe_line_correl}, together with the
soft and hard X-ray light curves. 

The shape of the EW variation is broadly similar to the profile at
$\Phi_{35}$=0.17 (R02), although that observation shows a more
`top-hat' profile. Consistent with R02, we find in the main-on state
that both the line normalization and the EW follow the soft X-ray
light curve and are anti-correlated with the hard, non-thermal flux
emitted above 2~keV.

\subsection{Variation of Fe complex over the orbital period}
\label{fe_orb}

\begin{figure}
\begin{center}
\setlength{\unitlength}{1cm}
\begin{picture}(8,5.)
\put(9.8,-1){\includegraphics{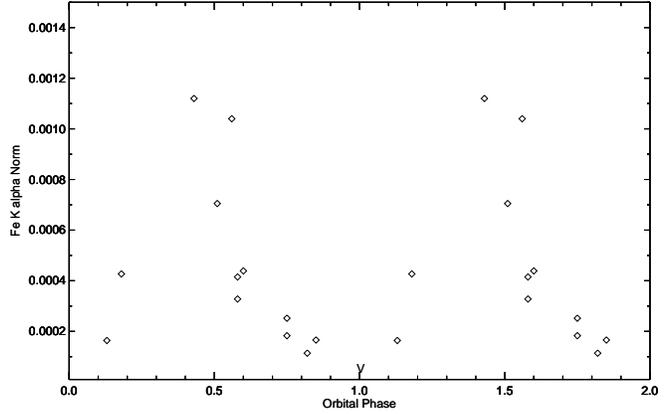}}
\end{picture}
\end{center}
\caption{The Fe 6.4keV K$\alpha$ line normalisation as a function of
the orbital phase, for all observations taken outside the
main-on.  Data have been taken with EPIC~PN. Only an upper limit has been 
obtained from the observation
taken during the eclipse (arrow).}
\label{fe_line_orb} 
\end{figure}

Further information about the origin of the Fe K$\alpha$ 6.4keV line
can be obtained by observing its variation over the orbital period. This
is shown in Figure~\ref{fe_line_orb}, for all observations taken
outside the main-on. The average line normalization is $2-4 \times
10^{-4}$ counts/cm$^2$/s, with larger peaks at $\sim 10^{-3}$
counts/cm$^2$/s and a dip at $\phi_{binary}=0.43-0.56$. (The line
normalisation during the main-on state is greater by a factor if
$>$5-10 compared to the low state). The line is not detected during
the middle of the eclipse, at which point the upper limit on the line
normalisation is $3.7\times10^{-5}$ counts/cm$^2$/s.

In the previous section we showed that during the main-on state the
line normalisation phased on the spin period is correlated with the
soft X-ray light curve suggesting a common origin for the thermal soft
X-ray emission and the 6.4keV line. We now find that in the low state
the line normalisation is correlated with the orbital period. We go
onto explore this further in \S \ref{fe_line_discuss}.

\section{Discussion}
\label{disc}

We have reported the results based on EPIC and OM data for a set of
{\xmm} observations of Her~X-1. Consistent with past reports, we find
that the spin modulation of the neutron star is most prominent during
the main-on and short-on state, becoming not significant during the
low state. These data sets allow us to follow the change in the
relative phasing of thermal and non-thermal emission and the change in
the pulse profile over the beat period. The large collective area of
{\xmm} and its good response down to soft X-rays also revealed a complex 
substructure in the light curve in the softest and less
well studied energy band.

Data taken during the main on, at $\Phi_{35} = 0.02$, show a common
phase modulation between the fluorescent Fe K$\alpha$ line and the
soft emission, confirming the results found by R02 at $\Phi_{35} =
0.17$. Moreover, for the very first time in CCD non grating spectra, 
we report evidence for a second Fe line, with central energy
near $\sim 7$~keV, which is only visible outside the main-on state. We
now proceed by discussing some physical interpretations for our
results.

\subsection{The spin-resolved light curves: evidence for
structure in soft X-rays ($0.3-1$~keV)}

The complex, multipeaked structure of the pulse profile of Her~X-1
above $\sim 1$~keV and its evolution pattern over the beat cycle is
the subject of several studies.  The basic features of the pulse
profile are nominally explained by the oblique rotator model (Lamb,
Pethick \& Pines 1973), but there are still many unresolved questions
regarding the details of the mechanism, such as whether the radiation
pattern is best described by a superposition of `fan' and `pencil'
beams and whether the obscuring material that modifies the visibility of
the direct beams from the neutron star contributes to the pulse
formation. Particularly successful in explaining the pattern observed
by {\it RXTE} and {\it Ginga} is the accretion column model by Scott
et al.~(2000). This model invokes the successive obscuration of a
direct beam (that originates from the polar caps of the neutron star),
two fan beams focussed in the antipodal direction (possibly due to
cyclotron backscattering) and a more extended scattered halo. As the
main-on progresses, the inner edge of the accretion disk cover first
one of the fan beam components, then the direct pencil beam and then
the second fan beam component.

In contrast, the pulsation detected below $\sim 1$~keV using {\sl
EXOSAT} (\"{O}gelmen \& Tr\"{u}mper~1988), {\sl ROSAT}
(Mavromatakis~1993), and {\sl BeppoSax} (Oosterbroek et al.~1997,
2000, 2001) is broad and quasi-sinusoidal. This emission is thermal
and probably originates in the inner edge of the disk, that partially
intercepts and reprocesses the hard X-rays from the neutron star.
{\xmm} data taken during the main-on have an excellent signal to noise
ratio and allow us to resolve a substructure in the
soft X-ray light curve. The 0.3-0.7~keV pulse profile observed at
$\Phi_{35}$=0.02 clearly shows the presence of two separate pulses, at
$\phi_{spin} \sim 0.0-0.1$ and $\phi_{spin} \sim 0.3-0.4$ in
Figure~\ref{spin_hard} (the value of the absolute spin phase is
arbitrary).  Similar features are also evident in data taken later in
the beat cycle: at $\Phi_{35}$=0.17 the pulse profile shows two maxima
(at $\phi_{spin} \sim 0.7$ and $\phi_{spin} \sim 0.9$ in
Figure~\ref{spin_hard}) and a small ``notch'' like feature preceding
the main pulse (at $\phi_{spin}$=0.4 in
Figure~\ref{spin_hard}). During the short on ($\Phi_{35}$=0.60), the
soft X-ray modulation is broader, more sinusoidal, and exhibits a complex 
substructure in the leading edge of the pulse followed
by a faster decay. 

Whether these features reflect an intrinsic
complexity in the thermal and geometrical properties of the
reprocessing region or keep memory of the structure of the
illuminating beam(s) is unclear. However, it is worth noticing that
during all three beat phases with bright emission, the features
detected in the soft light curves qualitatively resemble those of the
corresponding hard band. The two maxima observed in the softer band
during the main-on ($\Phi_{35}$=0.02, 0.17) have a phase separation
similar to that of the two main peaks observed in the 2-10~keV range,
i.e. $\Delta \phi_{spin} \sim 0.2-0.3$.  Similarly, the small
``notch'' preceeding the main pulse at $\Phi_{35}$=0.17 may be
associated with the atypical pulse observed at higher energies (both
precede the main peak by $\Delta \phi_{spin} \sim 0.25-0.3$). If these
associations are correct, i.e. if at least some of the main
characteristics of the neutron star beam are not ``washed out' during
the reprocessing of hard X-rays, this means that the radiative time
scale in the inner edge of the disk is much smaller than the photon
travel time.  Moreover, the reprocessing region cannot be too extended
in size. This may be explained by a high inclination angle of the
disk and/or a highly warped inner edge, in which case only a fraction
of the inner region of the disk intercepts the neutron star beam.
Alternatively, the reprocessing of the neutron star beam may be
dominated by a shocked, optically thick hot spot instead of being
distributed over the entire inner edge. A similar situation is
observed in SW Sex stars (see e.g. Dhillon et al.~1997, Groot et
al.~2000 and their Figure~6), where the hot spot region is formed
through a shock that occurs along the stream trajectory, but well
inside the accretion disk.

\subsection{The pulse period and the change in the phase shift between
soft and hard emission}

Given the complexity of the source, pulse-phase spectroscopy is of
paramount importance in disentangling the different spectral
components observed in Her X-1. Past observations using {\sl Einstein}
(McGray et al.~1982) and {\sl BeppoSax} (Oosterbroek at al. 1997,
2000) showed that, during the main-on state, the maximum of the
thermal component and the power-law components are shifted by $\sim
250^\circ$. The departure from a phase shift of $180^\circ$ is usually
associated with a disk having a tilt angle, under the assumption that
the soft X-rays originate in the layers of the disk that intercept and
reprocess the neutron star beam. The first evidence for a change in
the pulse difference, as measured at three different $\Phi_{35}$ using
{\xmm}, has been reported by R02. R02 found that during the main-on
state ($\Phi_{35}$=0.17) the soft and the hard X-ray pulse profiles
were strongly anti-correlated and that the phase shift between the
soft and hard maxima was $\sim150^{\circ}$. This was in contrast to
previous observations made during the main-on. Moreover, the phase
shift between these components was significantly different in the
{\xmm} observations performed at $\Phi_{35}$=0.26 and 0.60. This led
to the suggestion that, during the first three exposures, the tilt
angle of the accretion disk was changing substantially (R02). In
contrast with past observations (Oosterbroek et al.~2000), R02 did not
detect a symmetric state at $\Phi_{35} \sim 0, 0.5$. If the
reprocessing region is located at or near the inner region of the
disk, such lack of symmetry may indicate that the disk is strongly
warped (Heemskerk \& van Paradijs 1989). In order to further monitor 
the phase shift over the beat cycle, we presented a new measure of the
pulse difference, taken during the main-on ($\Phi_{35} = 0.02$). We
find that the trend detected by R02 is confirmed, and the change in
the tilt angle appears to be smooth and continuous over the beat
cycle.

Possible recent changes in the structure of the inner layers of the disk
have been also discussed by Oosterbroek et al.~(2001), who detected a
rapid spin down in Her~X-1 following an anomalous low state. This spin
down was not accompanied by a significant simultaneous change in the mass
accretion rate (Oosterbroek et al.~2001). Instead, it was more plausibly
associated with changes in the structure of the outer magnetosphere of the
neutron star and/or in the inner disk layers. These observations impose
important constraints on any theory modeling the accretion torques and
phenomena related to the inner edge of the disk in Her~X-1.

\subsection{The Fe K$\alpha$ line}
\label{fe_line_discuss}

The strong emission line at $\sim 6.4$~keV is detected in all {\xmm}\/
observations, with larger broadening and normalization during the
main-on. 
Past reports of Fe features in the spectra of Her~X-1 include an {\sl
ASCA} exposure (Endo et al.~2000) and a set of 63 {\sl Ginga}
observations analyzed by Leahy et al.~(2001).  The good coverage of
the beat period allowed Leahy et al.~(2001) to undertake a systematic
study of the line variation along the precession cycle. They found a
broad feature near $\sim 6.5$~keV, with the EW varying considerably
between main on, short on and low state. The mean values for each
state are 0.48~keV, 0.66~keV, and $\sim 1.37$~keV, respectively. Similar 
correlations are not evident in the {\xmm}\/ data, where instead 
an additional Fe XXVI line, possibly a coronal signature, is
detected during the low states (see \S~\ref{7keV}).

Nevertheless, we find evidence for a significant variation of the
line energy over the 35~day period: the Fe line emission originates in
near neutral Fe (Fe XIV or colder) in the low and short-on state
observations, whereas in the main-on the observed Fe K$\alpha$ centroid
energies ($6.65 \pm 0.1$~keV and $6.50 \pm 0.02$~keV as measured using
PN at $\Phi_{35}=0.02$ and 0.17) correspond to Fe XX-Fe XXI (Palmeri
et al.~2003). 
The line centroids observed using the EPIC PN deviate by $4 \sigma$
from the 6.40~keV neutral value. This has been already noticed by R02,
who suggested three possible explanations for both the line broadening
and the centroid displacement: 

\begin{enumerate}
{\item an array of Fe K$\alpha$ fluorescence lines
exists for a variety of charge states of Fe (anything from Fe I-Fe
XIII to Fe XXIII);}

{\item 
Comptonization from a hot corona with a
significant optical depth for a narrower range of charge states
centered around Fe XX. Similar line broadening has been observed in
some low mass X-ray binaries with ASCA (Asai et al.~2000), and Her~X-1
spectra exhibit one of the larger equivalent widths within the class;} 

{\item Keplerian motion: if this is responsable for the line broadening, 
the Keplerian velocity measured
at $\Phi_{35}=0.02$ and 0.17 is $\sim 15500$ and $\sim 13000$ km/sec,
respectively. This gives a radial distance of $\sim 2-3 \times
10^8$~cm (for a neutron star of 1.4 solar masses), which is close to
the magnetospheric radius for a magnetic field of $\sim 10^{12}$~G.} 

\end{enumerate}

In addition to these possibilities, we also notice that the region
responsible for the Fe K$\alpha$ line emission is likely to be different
for lines observed at different beat phases. The Fe lines have
different energies, flux, and broadening in the main-on compared to the
low state. While data taken during the
main on clearly suggest a correlation between the fluorescent Fe K$\alpha$
line and the soft X-ray emission, the same is not explicitly evident in
data taken during the low state. At such phases, instead, the line is a
factor $>5$ weaker and is clearly modulated with the orbital period. 

The correlation between the UV and the neutral Fe K point to a common 
origin with the fast rise UVW1 emission. Still, it is hard to 
distinguish the contributions from the companion and the disk to the Fe K 
line emission. A large contribution from the companion 
would explain why the Fe K line is
so strong in the low state of Her X-1 compared to other accretion disk 
sources: in Her X-1 the companion is much larger (2.3 \Msun) than 
for other LMXBs. However, due to the fast rise of the Fe K flux with the  
orbital phase, the disk origin scenario is still strong.

A possible contribution to the Fe K$\alpha$  emission may arise from 
relatively cold
material in an accretion disk wind, such as commonly observed in 
cataclysmic
variables (see e.g. Drew~1997). Although no
consensus has been reached in the case of low-mass X-ray
binaries, in the case of Her X-1 evidence of outflowing gas, possibly
a wind or material ablated from HZ Her, has been reported by Anderson
et al. (1996) and Boroson et al. (2001). UV lines observed with the
FOS and STIS spectrographs on the {\it Hubble Space Telescope} persist
even in the middle of the eclipse, when the X-ray heated atmosphere of
the normal star and the accretion disk should be entirely hidden from
view. The velocities inferred from the broadening of the N~V lines are
not constant over a time scale of $\sim 1$~yr, while a comparison
between observations taken at $\phi_{binary}$ =0.10, 0.21 and 0.29
show they are stable over the orbital phase.  Between these phases,
the velocity of the neutron star is expected to change by $\approx
60$~km/s. In this respect we note that the Fe K$\alpha$ line was not been 
detected by {\xmm 
\/} during the middle of the eclipse, and the 
upper limit on the line flux is
$\sim10$ or less of that measured outside the eclipse. This in turn
sets an upper limit on a potential contribution to the emission of this
line by a wind or some sort of circumstellar material. There is no Doppler 
signature of a wind in the HETG spectrum of the Fe K line (Jimenez-Garate 
et al.~2003), and the wind 
would have to be enclosed by the Roche lobe due to the absence of Fe K in 
eclipse. 

On the other hand, taken as a body the data reported here suggest a
complex origin for the overall emission of the Fe K$\alpha$ line. To our
knowledge, a complex of lines which include all ionization states from Fe
XVIII to XXIV Fe K$\alpha$ has not been observed in any astrophysical
source. This may still indicate an outflow of relatively cold gas or some
complex dynamics in the disk/magnetosphere interface. 
Such phenomena
should be time-dependent and may be monitored in the future using 
Astro-E2.

\subsection{The $\sim7$~keV Fe line}
\label{7keV}

Our observations have revealed an Fe~XXVI line at $\sim7$~keV, detected
during low-state. The line is not detected during the main-on when the
X-ray spectrum is dominated by the strong continuum emission. This
feature has not been detected by Leahy et al.~(2001) using {\sl Ginga}
data.  However, {\sl Ginga} has a relatively low spectral resolution
around these energies (18 percent at 6~keV). {\sl ASCA} observed the
source with better spectral resolution (2 percent at 5.9~keV using the
SIS detectors) during an early main on state, and based on these data
Endo et al.~(2000) reported the first resolution of the broad feature
into two components, at $6.41$~keV and $6.73$~keV.  However, this
double structure has not been detected in other observations taken
during the main-on nor is it present in {\xmm} data at similar phases.

The EPIC detectors on board {\xmm} have a spectral resolution similar
to {\sl ASCA} at 6~keV. However, {\xmm} has a much greater effective
area allowing lines to be detected with a greater confidence, in
particular during the states of low intensity of the binary
system. The feature at $\sim 7$~keV has been observed by {\xmm} for
the first time over several 35~day phases. Also, it has been confirmed
by a {\sl Chandra} HETGS observation of the source (the only one made
during the low state) taken at $\Phi_{35}= 0.44-0.46$ (Jimenez-Garate et 
al.~2003). The feature
cannot be produced by fluorescence, and it is more likely to be a
Fe~XXVI line originating in widely extended photo-ionized plasma. On
the other hand, RGS data taken during the low and short-on states also
show the presence of photo-ionized gas (Jimenez-Garate et al.~2002).
Grating spectra exhibit several narrow recombination emission lines,
the most prominent being C~VI, N~VI, N~VII, O~VII, O~VIII and
Ne~IX. The line ratio $G=(f+i)/r$, as computed for all the helium-like
ion complexes, is $G \simeq 4$, which indicates that photoionization
is the dominant mechanism.\footnote{Here $i$, $r$ and $f$ denote the
intercombination line blend, the resonance line and the forbidden
line; for collisionally ionized gases it is $G < 1$ (Liedahl 1999,
Porquet \& Dubau 2000).}  Moreover, RGS spectra shows two radiative
recombination continua of O~VII and N~VII, consistent with a low
temperature of the emitting plasma (30000~K~$<T<$~60000~K,
Jimenez-Garate et al.~2002), which furtherly validate the
photoionization model. None of these features is detected during the
main-on state (where instead there might be evidence for a broad O VII i 
line, and perhaps O
VIII Ly alpha). The recombination X-ray line emission are not likely to
originate in HZ Her, due to the absence of UV induced photoexcitation
signatures in the He-like triplets (observed with HETGS,
Jimenez-Garate private comm.).

Instead, as suggested by Jimenez-Garate et al.~(2002), an extended,
photoionized accretion disk atmosphere and corona may be responsable
for such features. Jimenez-Garate et al.~(2001) computed the spectra
of the atmospheric layers of a Shakura-Sunyaev accretion disk,
illuminated by a central X-ray continuum.  They found that, under
these conditions, the disk develops both an extended corona which is
kept hot at the Compton temperature, and a more compact, colder, X-ray
recombination-emitting atmospheric layer (see Figure~9 in
Jimenez-Garate et al.~2002). The parameters of the compact atmospheric
layer are consistent with the constraints to the narrow line emitting
region, as derived from spectroscopic analysis and modeling of the RGS
features.

\begin{figure} 
\begin{center} 
\setlength{\unitlength}{1cm}
\begin{picture}(8,5) 
\put(-0.1,0.){\includegraphics{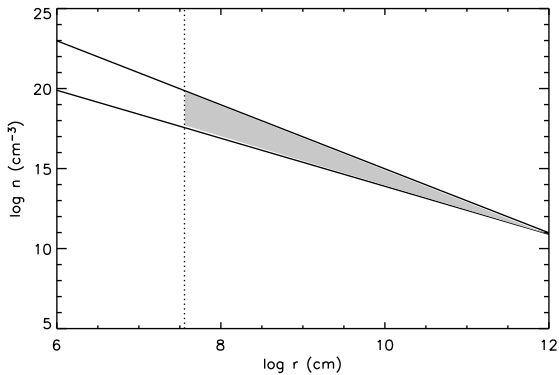}} 
\end{picture} 
\end{center} 
\caption{Electron density  plotted versus the size of the region enclosing 
the Fe XXVI line emitting plasma. The
three lines are the constraints discussed in the text and the shadowed
region is allowed.} 
\label{iron_reg} 
\end{figure}

Interestingly, the Fe XXVI line detected by {\xmm}\/ may be a
signature of the {\it hottest} external layers of the disk corona,
which are located above the recombination-emitting layers. In order to
detect a line at $\sim 7$~keV, an ionization parameter $\log \xi \geq
3.3$ is required (Kallman and McCray~1982, their Fig.~1b).  Using the
flux of power law component reported by R02 (i.e. $3.65 
\times 10^{-8}$ 
erg/cm$^2$/s at $\Phi_{35}=0.17$) and a source distance of
$6.6$~kpc, the luminosity of the direct beam is $L_x=1.9 \times
10^{38}$~erg/s. This gives
\begin{equation} n r^2 \equiv \frac{L_x}{\xi} < 9.7 \times 10^{34}
\, {\rm cm}^{-1} \, ,
\end{equation} 
where $n$ is the particle density and $r$ is the size of
the region enclosing the Fe XXVI line emitting plasma. A second
constraint can be derived from the emission measure. Running a detailed 
multi-temperature plasma model in XSTAR gives an emission measure of $n^2 
V> 6.1 \times 10^{57}$~cm$^{-3}$ for the 7~keV line, for a flux of 
$2 \times 10^{-4}$ photons/cm$^{2}$/s (the emission measure scales 
linearly with 
the line flux, and the latter varies between $1-2 \times10^{-4}$ 
photons/cm$^{2}$, see Figure~\ref{fe_line})\footnote{The value of the 
emission measure is consistent with 
calculations by Kallman et al.~(2003), who found $10^{57}$ cm$^{-3}$ 
for an accretion disk corona at 7 kpc and a Fe XXVI line with a roughly 
similar flux level.}
. The average
temperature, computed by using a main-on custom ionization spectrum in
XSTAR as well as HULLAC recombination emissivities (D. Liedahl,
private comm.), is $kT \sim 580$~eV.  Moreover, if the line emitting
plasma is located outside the Alfv\'en shell, $r>3.6 \times
10^7$~cm. These three constraints define an allowed region in the
plane $n-r$, which is shown in Figure~\ref{iron_reg}.  The density of
the hot coronal layers in the disk models by Jimenez-Garate et
al.~(2001) is $10^{15}-10^{16}$~cm$^{-3}$ if the radius is $r \sim
10^8-10^{10}$~cm, therefore within the limits shown in the figure.

In summary, the variability of the Her~X-1 spectrum lends support to
the precession of the accretion disk.  The evidence for the disk
identification relies on the modeled structure and spectra from a
photoionized disk (Jimenez-Garate et al.~2001), which is in agreement
with the limit set spectroscopically on the density of the low-energy
lines emitting region.  Interestingly, the same model naturally
provides a candidate for the region emitting the Fe~XXVI line and
again the computed values of the density agree with the constraints
inferred from the line parameters.  Taken together, these facts
strengthen the interpretation of the low state emission in term of an
extended source and open the exciting possibility to monitor
spectroscopically the different atmospheric components of the disk
during the transition from the low to the high state.

\section{Acknowledgments}

We would like to thank the referee, Arvind Parmar, for a careful
reading of the manuscript. Based on observations obtained with
XMM-Newton, an ESA science mission with instruments and contributions
directly funded by ESA Member States and the USA (NASA). This paper
makes use of quick-look results provided by the ASM/RXTE team whom we
thank.


\begin{thebibliography}{}

\bibitem[]{as00} Asai, K., Dotani, T., Nagase, F., \& Mitsuda, K., 2000,
ApJS, 131, 571
\bibitem[]{bay93} Baykal, A., Boynton, P. E., Deeter, J. E., \& Scott, D.
M., 1993, MNRAS, 267, 347
\bibitem[]{bo01} Boroson, B., Kallman, T., \& Vrtilek, S.D., 2001,
ApJ, 562, 925
\bibitem[]{c79} Cash, W., 1979, ApJ, 228, 939
\bibitem[1994]{ch1994} Choi, C.S., Nagase, F., Makino, F., Dotani, T.,
Kitamoto, S., \& Takahama, S. 1994, ApJ, 437, 449
\bibitem[]{co00} Coburn, W., Heindl, W. A., Wilms, J., Gruber, D. E.,
Staubert, R., Rothschild, R. E., Postnov, K. A., Shakura, N., Risse, P.,
Kreykenbohm, I., \& Pelling, M. R., 2000, ApJ, 543, 351
\bibitem[1998]{de1998} Deeter, J.E., Scott, D. M.,
Boynton, P. E., Miyamoto, S., Kitamoto, S., Takahama, S., \& Nagasem F.
1998, ApJ, 502, 802
\bibitem[1997]{dhi97} Dhillon, V. S., Marsh, T. R., \& Jones, D. H. P.,
1997, MNRAS, 291, 694
\bibitem[1998]{d084} Dotani, T., Kii, T., Nagase, F., Makishima, K.,
Ohashi, T., Sakao, T., Koyama, K., \& Tuohy, I., 1989, PASJ< 41, 427
\bibitem[]{dr97} Drew, J., 1997, in IAU colloquium 197, Accretion
Phenomena and Related Outflows, ed. D.T. Wickramasinghe, L.
Ferrario, \& G.V. Bicknell (ASP Conf. Ser., 121; San
Francisco:ASP), 465
\bibitem[2000]{en2000} Endo, T., Nagase, F., \& Mihara, T. 2000, PASJ, 52,
223
\bibitem[]{gro00} Groot, P. J., Rutten, R. G. M., \& van Paradijs,
J., 2000, NewAR, 44, 137
\bibitem[]{ger76} Gerend, D., \& Boynton, P., 1976, Apj, 209, 562
\bibitem[]{hee89} Heemskerk, M. H. M., \& van Paradijs, J., 1989, A\&A,
223, 154
\bibitem[]{mario01} Jimenez-Garate, M. A., Raymond, J. C., Liedahl, D. A.,
\&  Hailey, C. J., 2001, ApJ, 558, 448
Hailey, C. J., den Herder,
\bibitem[]{mario02} Jimenez-Garate, M. A., Hailey, C. J., den Herder,
J.-W.,
Zane, S., \& Ramsay, G., 2002, ApJ, 578, 391
\bibitem[]{mario03} Jimenez-Garate, M.A., Raymond, J. C., Liedahl, D. A., 
\& Hailey, C.J., 2003, ApJ submitted 
\bibitem[]{kall82} Kallman, T.R., \& McCray, R., 1982, ApJS, 50, 263
\bibitem[]{kall03} Kallman, T.R., Angelini, L., Boroson, B., \& Cottam, 
J., 2003, ApJ, 583, 861
\bibitem[]{lamb73} Lamb, F.K., Pethick, C.J., \& Pines, D., 1973, ApJ,
184, 271
\bibitem[]{le01} Leahy, D.A., 2001, ApJ, 547, 449
\bibitem[]{lev91} Levine, A., Rappaport, S., Putney, A., Corbet, R., \&
Nagase, F., 1991, ApJ, 381, 101
\bibitem[]{li99} Liedahl, D.A., Wojdowski, P., Jimenez-Garate, M.A., \&
Sako, M., 2001, in ASP Conf. Ser. 247, Spectroscopic Challenges of
Photoionized Plasmas, ed. G. Ferland \& D. W. Savin (San Francisco: ASP),
447
\bibitem[]{mas01} Mason, K. O., Breeveld, A., Much, R., Carter, M.,
Cordova, F. A., Cropper,
M. S., Fordham, J., Huckle, H., Ho, C., Kawakami, H., and 9 coauthors,
2001, A\&A, 365, L36
\bibitem[]{mav93} Mavromatakis, F., 1993, A\&A, 273,
147 \bibitem[1982]{mc1992} McCray, R.A., Shull, J. M.,
 Boynton, P. E., Deeter, J. E., Holt, S. S., \& White, N. E. 1982, ApJ,
262,
301
\bibitem[1991]{mi1991} Mihara, T., Ohashi, T., Makishima, K., Nagase, F.,
 Kitamoto, S., \& Koyama, K., 1991, PASJ, 43, 501
\bibitem[]{oge98} Ogelmen, H., \& Tr\"{u}mper, J., 1988, MmSAI, 59, 169
\bibitem[1997]{oo1997} Oosterbroek, T.
Parmar, A. N., Martin, D. D. E., \& Lammers, U., 1997, A\&A, 327, 215
\bibitem[2000]{oo2000} Oosterbroek, T., Parmar, A. N., Dal Fiume, D.,
Orlandini, M., Santangelo, A., Del Sordo, S., \& Segreto, A., 2000,
A\&A, 353, 575
\bibitem[2001]{oo2001} Oosterbroek, T., Parmar, A. N., Orlandini, M.,
Segreto, A., Santangelo, A., \& Del Sordo, S., 2001, A\&A, 375, 922
\bibitem[1989]{pa89} Parmar, A. N., White, N.E., \& Stella, L., 1989, ApJ,
338, 373
\bibitem[1999]{pa99} Parmar, A. N., Oosterbroek, T., dal Fiume, D.,
Orlandini, M., Santangelo, A.,  Segreto, A., \& del Sordo, S., 1999, A\&A,
350, L5
\bibitem[2003]{pa03} Palmeri, P., Mendoza, C., Kallman, T. R. \& Bautista,
M. A., 2003, A\&A, 403, 1175
\bibitem[2000]{po00} Porquet, D. \& Dubau, J., 2000, A\&AS, 143, 495
\bibitem[]{gavin02} Ramsay G., Zane S., Jimenez-Garate M. A., den Herder
J.-W., \& Hailey C. J., 2002, MNRAS, 337, 1185 (R02)
\bibitem[]{scott00} Scott, M. D., Leahy, D. A., \& Wilson, R. B., 2000,
ApJ, 539, 392
\bibitem[]{sti01} Still M., O'Brien K., Horne K., Hudson D., Boroson B.,
Vrtilek S. D., Quaintrell H., \& Fiedler H., 2001, ApJ, 553, 776
\bibitem[2001]{st2001} Str\"{u}der, L., Briel, U., Dennerl, K., Hartmann,
R., Kendziorra, E.,
Meidinger, N., Pfeffermann, E., Reppin, C., Aschenbach, B., Bornemann, W.,
and 48 coauthors, 2001, A\&A, 365, L18
\bibitem[1998]{su98} Sunyaev, R. A., Gilfanov, M. R.,
Churazov, E. M., Loznikov, V. M., Efremov, V. V., Kaniovskii, A. S.,
Kuznetsov, A. V., Melioranskii, A. S., Voges, W., Pietsch, W., and 11
coauthors, 1988, SvAL, 14, 416
\bibitem[2001]{tu2001} Turner, M. J. L., Abbey, A., Arnaud, M., Balasini,
M., Barbera, M.,
Belsole, E., Bennie, P. J., Bernard, J. P., Bignami, G. F., Boer, M., and
53 coauthors, 2001, A\&A, 365, L27
\bibitem[]{}Watanabe, S., Sako, M., Ishida, M., Ishisaki, Y., Kahn, 
S.M., Kohmura, T., Morita, U., Nagase, F., Paerels, F., \&  
Takahashi, T., 2003, ApJ, 597, L37 
\bibitem[1996]{vrt96} Vrtilek, S. D., \& Cheng, F.H., 1996, ApJ, 465, 915


\end{thebibliography}
\end{document}